\documentclass[fleqn,usenatbib]{mnras}

\usepackage{newtxtext,newtxmath}

\newcommand{\Ha}{$\rm{H} \alpha$}

\newcommand{\PVdblt}{{\rm P}\kern 0.1em{\sc v}~$\lambda\lambda 1117, 1128$}
\newcommand{\CaIIdblt}{{\rm Ca}\kern 0.1em{\sc ii}~$\lambda\lambda 3934, 3969$}
\newcommand{\AlIIIdblt}{{\rm Al}\kern 0.1em{\sc iv}~$\lambda\lambda 1855, 1863$}
\newcommand{\CIVdblt}{{\rm C}\kern 0.1em{\sc iv}~$\lambda\lambda 1548, 1550$}
\newcommand{\MgIIdblt}{{\rm Mg}\kern 0.1em{\sc ii}~$\lambda\lambda 2796, 2803$}
\newcommand{\NVdblt}{{\rm N}\kern 0.1em{\sc v}~$\lambda\lambda 1238, 1242$}  
\newcommand{\SVIdblt}{{\rm S}\kern 0.1em{\sc vi}~$\lambda\lambda 933, 944$} 
\newcommand{\OVIdblt}{{\rm O}\kern 0.1em{\sc vi}~$\lambda\lambda 1031, 1037$} 
\newcommand{\SiIIdblt}{{\rm Si}\kern 0.1em{\sc ii}~$\lambda\lambda 1190, 1193$} 
\newcommand{\SiIVdblt}{{\rm Si}\kern 0.1em{\sc iv}~$\lambda\lambda 1393, 1402$} 
\newcommand{\PV}{\hbox{{\rm P}\kern 0.1em{\sc v}}}
\newcommand{\AlI}{\hbox{{\rm Al}\kern 0.1em{\sc i}}}
\newcommand{\AlII}{\hbox{{\rm Al}\kern 0.1em{\sc ii}}}
\newcommand{\AlIII}{{\hbox{\rm Al}\kern 0.1em{\sc iii}}}
\newcommand{\CaII}{\hbox{{\rm Ca}\kern 0.1em{\sc ii}}}
\newcommand{\CII}{\hbox{{\rm C}\kern 0.1em{\sc ii}}}
\newcommand{\CIIe}{\hbox{{\rm C$^{\ast}$}\kern 0.1em{\sc ii}}}
\newcommand{\CIII}{\hbox{{\rm C}\kern 0.1em{\sc iii}}}
\newcommand{\CIV}{\hbox{{\rm C}\kern 0.1em{\sc iv}}}
\newcommand{\CV}{\hbox{{\rm C}\kern 0.1em{\sc v}}}
\newcommand{\HI}{\hbox{{\rm H}\kern 0.1em{\sc i}}}
\newcommand{\HII}{\hbox{{\rm H}\kern 0.1em{\sc ii}}}
\newcommand{\Lya}{\hbox{{\rm Ly}\kern 0.1em$\alpha$}}
\newcommand{\Lyb}{\hbox{{\rm Ly}\kern 0.1em$\beta$}}
\newcommand{\Lyg}{\hbox{{\rm Ly}\kern 0.1em$\gamma$}}
\newcommand{\Lyd}{\hbox{{\rm Ly}\kern 0.1em$\delta$}}
\newcommand{\Lye}{\hbox{{\rm Ly}\kern 0.1em$\epsilon$}}
\newcommand{\Lyphi}{\hbox{{\rm Ly}\kern 0.1em$\phi$}}
\newcommand{\Lyfive}{\hbox{{\rm Ly}\kern 0.1em$5$}}
\newcommand{\Lysix}{\hbox{{\rm Ly}\kern 0.1em$6$}}
\newcommand{\Lyseven}{\hbox{{\rm Ly}\kern 0.1em$7$}}
\newcommand{\Lyeight}{\hbox{{\rm Ly}\kern 0.1em$8$}}
\newcommand{\Lynine}{\hbox{{\rm Ly}\kern 0.1em$9$}}
\newcommand{\Lyten}{\hbox{{\rm Ly}\kern 0.1em$10$}}
\newcommand{\Lyeleven}{\hbox{{\rm Ly}\kern 0.1em$11$}}
\newcommand{\HeI}{\hbox{{\rm He}\kern 0.1em{\sc i}}}
\newcommand{\HeII}{\hbox{{\rm He}\kern 0.1em{\sc ii}}}
\newcommand{\FeI}{\hbox{{\rm Fe}\kern 0.1em{\sc i}}}
\newcommand{\FeII}{\hbox{{\rm Fe}\kern 0.1em{\sc ii}}}
\newcommand{\FeIII}{\hbox{{\rm Fe}\kern 0.1em{\sc iii}}}
\newcommand{\MnII}{\hbox{{\rm Mn}\kern 0.1em{\sc ii}}}
\newcommand{\MgI}{\hbox{{\rm Mg}\kern 0.1em{\sc i}}}
\newcommand{\MgIb}{\hbox{{\rm Mg}\kern 0.1em{\sc i}}\kern 0.05em{\rm b}}
\newcommand{\MgII}{\hbox{{\rm Mg}\kern 0.1em{\sc ii}}}
\newcommand{\MgIII}{\hbox{{\rm Mg}\kern 0.1em{\sc iii}}}
\newcommand{\NI}{\hbox{{\rm N}\kern 0.1em{\sc i}}}
\newcommand{\NII}{\hbox{{\rm N}\kern 0.1em{\sc ii}}}
\newcommand{\NIII}{\hbox{{\rm N}\kern 0.1em{\sc iii}}}
\newcommand{\NV}{\hbox{{\rm N}\kern 0.1em{\sc v}}}
\newcommand{\OVI}{\hbox{{\rm O}\kern 0.1em{\sc vi}}}
\newcommand{\OI}{\hbox{{\rm O}\kern 0.1em{\sc i}}}
\newcommand{\OII}{\hbox{[{\rm O}\kern 0.1em{\sc ii}]}}
\newcommand{\OIII}{\hbox{[{\rm O}\kern 0.1em{\sc iii}]}}
\newcommand{\OIV}{\hbox{{\rm O}\kern 0.1em{\sc iv}]}}
\newcommand{\SI}{{\rm S}\kern 0.1em{\sc i}}
\newcommand{\SIV}{{\rm S}\kern 0.1em{\sc iv}}
\newcommand{\SVI}{{\rm S}\kern 0.1em{\sc vi}}
\newcommand{\SiI}{\hbox{{\rm Si}\kern 0.1em{\sc i}}}
\newcommand{\SiII}{\hbox{{\rm Si}\kern 0.1em{\sc ii}}}
\newcommand{\SiIII}{\hbox{{\rm Si}\kern 0.1em{\sc iii}}}
\newcommand{\SiIV}{\hbox{{\rm Si}\kern 0.1em{\sc iv}}}
\newcommand{\SII}{\hbox{{\rm S}\kern 0.1em{\sc ii}}}
\newcommand{\SIII}{\hbox{{\rm S}\kern 0.1em{\sc iii}}}
\newcommand{\NaI}{\hbox{{\rm Na}\kern 0.1em{\sc i}}}
\newcommand{\NaID}{\hbox{{\rm Na}\kern 0.1em{\sc i}}\kern 0.05em{\rm D}}
\newcommand{\TiII}{\hbox{{\rm Ti}\kern 0.1em{\sc ii}}}
\newcommand{\kms}{\hbox{~km~s$^{-1}$}}
\newcommand{\cmsq}{\hbox{cm$^{-2}$}}

\usepackage[T1]{fontenc}

\DeclareRobustCommand{\VAN}[3]{#2}
\let\VANthebibliography\thebibliography
\def\thebibliography{\DeclareRobustCommand{\VAN}[3]{##3}\VANthebibliography}

\usepackage{graphicx}	
\usepackage{amsmath}	
\usepackage{multirow}
\usepackage{setspace}
\usepackage{bigdelim}
\usepackage{bigstrut}
\usepackage{floatflt}	
\usepackage{wrapfig}
\usepackage{subfig}
\usepackage{multicol}
\usepackage[utf8]{inputenc}
\usepackage[flushleft]{threeparttable}

\defcitealias{muzahid15}{M15}

\title[Disentangling the multi-phase CGM]{Disentangling the multi-phase circumgalactic medium shared between a dwarf and a massive star-forming galaxy at z$\sim$0.4}

\author[Nateghi et al.]{
Hasti Nateghi,$^{1,2}$\thanks{E-mail: hnateghi@swin.edu.au}
Glenn G. Kacprzak$^{1,2}$,
Nikole M. Nielsen$^{1,2}$, 
Sowgat Muzahid$^{3}$, 
\newauthor Christopher W. Churchill$^{4}$,
Stephanie K. Pointon$^{2}$,
Jane C. Charlton$^{5}$\\
$^{1}$Centre for Astrophysics and Supercomputing, Swinburne University of Technology, Hawthorn, Victoria 3122, Australia\\
$^{2}$ARC Centre of Excellence for All Sky Astrophysics in 3  Dimensions (ASTRO 3D)\\
$^{3}$Leibniz-Institute for Astrophysics Potsdam (AIP), An der Sternwarte 16, D-14482 Potsdam, Germany\\
$^{4}$Department of Astronomy, New Mexico State University, Las Cruces, NM 88003, USA\\
$^{5}$Department of Astronomy and Astrophysics, The Pennsylvania State University, State College, PA 16801, USA\\
}

\date{Accepted 2020 November 10. Received 2020 November 09; in original form 2020 September 07}

\pubyear{2020}

\begin{document}
\label{firstpage}
\pagerange{\pageref{firstpage}--\pageref{lastpage}}
\maketitle

\begin{abstract}
The multi-phase circumgalactic medium (CGM) arises within the complex environment around a galaxy, or collection of galaxies, and possibly originates from a wide range of physical mechanisms. In this paper, we attempt to disentangle the origins of these multi-phase structures and present a detailed analysis of the quasar field Q0122$-$003 field using Keck/KCWI galaxy observations and {\it HST}/COS spectra probing the CGM. Our re-analysis of this field shows that there are two galaxies associated with the absorption. 
We have discovered a dwarf galaxy, G\_27kpc ($M_{\star}=10^{8.7}$~M$_{\odot}$), at  $z=0.39863$ that is 27~kpc from the quasar sightline. G\_27kpc is only $+21$~{\kms} from a more massive ($M_{\star}=10^{10.5}$~M$_{\odot}$) star-forming galaxy, G\_163kpc, at an impact parameter of 163~kpc.
While G\_163kpc is actively forming stars (${\rm SFR}=6.9$~M$_{\odot}$~yr$^{-1}$), G\_27kpc has a 
low star-formation rate (${\rm SFR}=0.08\pm0.03$~M$_{\odot}$~yr$^{-1}$) and star formation surface density ($\Sigma_{\rm SFR}=0.006$~M$_{\odot}$~kpc$^{-2}$~yr$^{-1}$), implying no active outflows. By comparing galaxy SFRs, kinematics, masses and distances from the quasar sightline to the absorption kinematics, column densities and metallicities, we have inferred the following: (1) Part of the low-ionization phase has a metallicity and kinematics consistent with being accreted onto G\_27kpc. (2) The remainder of the low ionization phase has metallicities and kinematics consistent with being intragroup gas being transferred from G\_27kpc to G\_163kpc. (3) The high ionization phase is consistent with being produced solely by outflows originating from the massive halo of G\_163kpc. Our results demonstrate the complex nature of the multi-phase CGM, especially around galaxy groups, and that detailed case-by-case studies are critical for disentangling its origins.
\end{abstract}




\section{Introduction}

The circumgalactic medium (CGM) is typically thought of as a gas reservoir that resides between the interstellar medium (ISM) and the virial radius of individual galaxies \citep[$\thickapprox 200$~kpc, e.g.,][]{Glenn2008,chen2010may,Tumlinson11,Tumlison17,Nikki13a_magiicat1}. The CGM is where inflowing gaseous material from the intergalactic medium (IGM) meets the galactic feedback environment, which includes outflows, recycled accretion, and tidally-stripped gas \citep{Tumlison17}.
Theoretical simulations predict that recycled feedback and wind/outflow processes play a prominent role in regulating the stellar mass and star formation rate (SFR) of galaxies \citep{springel2003,Oppenhimer08,Dave11}, where outflows are commonly observed around typical galaxies \citep[e.g.,][]{weiner2009,steidel2010,Glenn14,Rubin2014,hadi18,hadi18b,schroeter_19_megaflow3}.

Since the first detection of galaxies associated with intervening absorption systems found in the spectra of bright background quasars \citep{Bergon88,Bergon91,Steidel94}, studies have focused on the evolution and understanding of the CGM under the assumption that ordinarily only one galaxy is responsible for the absorption.
Based on the assumption that the CGM is associated with individual galaxies, the geometric distribution of the CGM has provided some insight into gas flows. The dependence of gas flows on galaxy inclination and azimuthal angle has been considered in many studies using the absorption line equivalent width measurements \citep{Bordoloi2011, Bouche2012, Glenn2012, Glenn2015dec, Lan2014}.  \citet{Bouche2012} and \citet{Glenn2012} found a bimodality in the azimuthal distribution of {\MgII} absorption systems around galaxies, where absorption prefers to exist along the projected major and minor axes of galaxies.  In addition, {\MgII} absorption tends to be much stronger along the minor axis of galaxies \citep{Bordoloi2011,Lan2014,Lan2018}. It has also been reported that the absorption velocity dispersions vary with azimuthal angle, which reflect signatures of outflows and accretion \citep{Nikki15_magii5}.

Investigating the relative kinematics between quasar absorption lines and their host galaxies enables us to determine the physical processes ongoing within the CGM. Several studies have shown that the CGM, as traced by {\MgII} absorption, co-rotates with galaxies out to large impact parameters along the projected major axis \citep{Steidel2002,Glenn2010,Glenn2011june,Glenn2017,Martin12,Rubin2012,Bouche13,Ho17,Lopez2020}, which provides strong evidence that the gas is accreting onto galaxies \citep{stewart2013,Stewart2017}. It has also been found that the CGM velocity structure can be well modelled by outflowing gas, which tends to be biconical \citep{Bordoloi2011,Bordoloi2014B,Bouche2012,Lan2014,Rubin2014}. With these models, outflow masses, rates and loading factors have also been obtained \citep{Bouche2012,Glenn14,Schroetter2015,Schroetter2016,Nikki2020_cosmicnoon}.

However, we know that the hypothesis that each galaxy has its own CGM is likely untrue, and it is still debated which type of environment is typically selected when finding CGM absorption in quasar spectra. \cite{schroeter_19_megaflow3} constructed a survey of 22 quasar lines of sight and found that $\sim80$ percent of their {\MgII} absorption systems corresponded to isolated galaxies with no companion within 100~kpc. In contrast, \citet{Aleks2020} found that $\sim38$ percent of their {\MgII} absorption systems were associated with isolated galaxies within 250~kpc projected distance from quasar sightline. 

We already know from observations of galaxies that the environment plays a prominent role in their evolution \citep[e.g.,][]{dressler1980}. Tidal stripping caused by mergers and interactions are able to remove large reservoirs of gas from the ISM and can lead to the quenching of star formation \citep[e.g.,][]{Cowie77,nulsen82,Cen99,Oppenhimer08,Lilly_2013}. Thus, environment can affect the SFRs, perturb the stable disk structure of galaxies \citep[e.g.,][]{Veilleux2005,poggianti2016}, and must have a significant impact on the CGM. Observations of cool {\HI} gas reveal different substructures as a consequence of galactic interactions in group environments such as tidal streaming gas and warped disks, as well as high-velocity clouds \citep[e.g.,][]{Fraternali2002,Chynoweth08,Mihos12,Wolf13}.

Simulations have shown that galaxy environment and mergers have an effect on the CGM. 
Using the Illustris simulations, \cite{Hani18} found that the covering fraction of high column density gas such as {\HI}, {\CIV} and {\OVI} increases before a major merger and this increase remains for billions of years after the merger.
In addition, the FIRE simulations have demonstrated that, at lower redshifts, intergalactic transfer may dominate the overall gas accretion in group environments \citep{Firesim2017}. These simulations suggest that the CGM of galaxies are not isolated and galaxies may also share, or have interacting, CGM. 

Recent studies have begun to concentrate on absorption-line systems associated with more than one galaxy \citep{Burchett2016,Bielby2017,Peroux17,Peroux2019,Pointon2017,Pointon2020,Nielsen2018_magiiVI,hadi18,Chen2019,Aleks2020,lehner2020AMIGA}. 
It has been found that there is no strong anti-correlation between the equivalent width of {\MgII} absorption and impact parameter for galaxy groups \citep{Chen2010dec,Nielsen2018_magiiVI}, which differs from the strong anti-correlation found for isolated galaxies \citep[e.g.][]{Steidel94, Glenn2008, Nikki2013b_magiicat2}. \cite{Bordoloi2011} found that {\MgII} absorption is stronger for group environments at fixed impact parameter, which was interpreted as being caused by a superposition of isolated galaxy halos within the group. \cite{Nielsen2018_magiiVI} also found statistically larger equivalent widths and covering fractions for galaxy groups. They determined that while the superposition model reproduces the equivalent widths, it severely over-predicts the velocity spread of the absorption. The authors concluded that the {\MgII} absorption arose from intragroup gas, further supporting the idea of a shared or combined CGM between galaxies. This is also supported by other studies that found that {\MgII} absorption likely arose from intragroup gas and/or tidal streams \citep{Whiting2006,Glenn2010b,Bielby2017,Aleks2020}.

Higher ionization gas, such as {\CIV}, has also been used to investigate group environments. \cite{Burchett2016} found an anti-correlation between the number of galaxies in their groups and the equivalent width of {\CIV} absorption. Interestingly, they did not detect {\CIV} in groups that include more than seven galaxies. It was also shown that {\OVI} absorption within galaxy groups exhibits lower column densities and lower velocity spreads when compared to isolated environments \citep{Stock13,Pointon2017,MasonNg2019}. This is consistent with cosmological simulations that show that since the virial temperature increases with halo mass, group environments ionize a larger fraction of oxygen to higher levels, which results in a lower ionization fraction of {\OVI} compared to isolated galaxies \citep{oppenheimer16}.
All of the aforementioned studies show that the multi-phase nature of the CGM in galaxy groups is complex, with different results found between low to high ionization levels. This could imply that the multi-phase gas associated with groups may arise from different physical mechanisms, and highlights the importance of examining a range of gas phases in galaxy groups. 

Given that environment plays a critical role in the properties of the CGM and the evolution of galaxies, it is imperative that we have a high spectroscopic completeness of galaxies in quasar fields.
 Using powerful integral field spectrographs like KCWI \citep{morrissey18_kcwi} and MUSE \citep{Bacon2006_muse} has efficiently provided us with the spectra of many sources in the field of view around quasars. At lower redshifts VLT/MUSE has been particularly effective in finding faint galaxies residing at the same redshift as a host-CGM galaxy that was previously thought to be isolated \citep{Peroux17,hadi18}. \citet{Peroux17} concluded that the {\MgII} absorption arose from tidal material within their galaxy group, while \citet{hadi18} associated the {\MgII} absorption to an individual galaxy within their group. These case-by-case studies are important for understanding the origins of the CGM within galaxy groups. Given the differing trends in line tracers with environment, multi-phase gas information could provide additional insight into the origins of this gas.


In \citet[][hereafter \citetalias{muzahid15}]{muzahid15} we identified an absorption-line system in Q0122$-$003 with a complex multiphase nature and a varied range in metallicities, which points to a possible complex origin of the CGM. The CGM was originally thought to originate within outflows from a distant massive star-forming galaxy. However, it remained difficult to assign all of the varied absorption-line properties to outflows of a single galaxy.  Motivated by the knowledge that galaxies do not reside in isolation and that a census of galaxies near the quasar was incomplete in this field, we use Keck/KCWI observations to identify additional galaxies that could be associated with the absorption. Our new KCWI data have allowed us to discover a new galaxy at the same redshift as the absorber and previously known galaxy, and the new galaxy is only $27$~kpc away from the quasar sightline.  The presence of this new low-mass ($M_{\star}= 10^{8.7}$~M$_{\odot}$) faint galaxy ($0.04L^{\star}_{B}$)  near the quasar sightline provides an alternate explanation for this complex multiphase absorbing gas. In this paper we use our KCWI data and UV and optical spectra of the quasar in an effort to disentangle the origin of the observed intervening gas.

The paper is organized as follows: 
In Section~\ref{sec:data} we describe the data and analysis. In Section~\ref{result} we present the results and discuss the properties of the new galaxy. We also summarize the results of \citetalias{muzahid15} for context and to provide the details required for our new analysis and interpretations. In Section~\ref{sec:discussion}, we discuss the origins of the low- and high-ionization phases and we present our concluding remarks in Section~\ref{sec:conclusion}. Throughout we adopt an ${\rm H}_{\rm 0}=70$~\kms~Mpc$^{-1}$, $\Omega_{\rm M}=0.3$, $\Omega_{\Lambda}=0.7$ cosmology and magnitudes are quoted using the AB system \citep{AB-mag}.


 
\section{Observations}
\label{sec:data}


In this Section, we present new KCWI observations and additional analysis of galaxies in the Q0122$-$003 field. For the benefit of the reader, we also summarise some of the data and analysis from  \citetalias{muzahid15}, which is critical for our analysis and interpretations. Our analysis of KCWI data provides new insights used to interpret the origins of the CGM along this sightline. 

\begin{figure*}
        \includegraphics[trim={0 2.5cm 0 2.5cm}, scale=0.65]{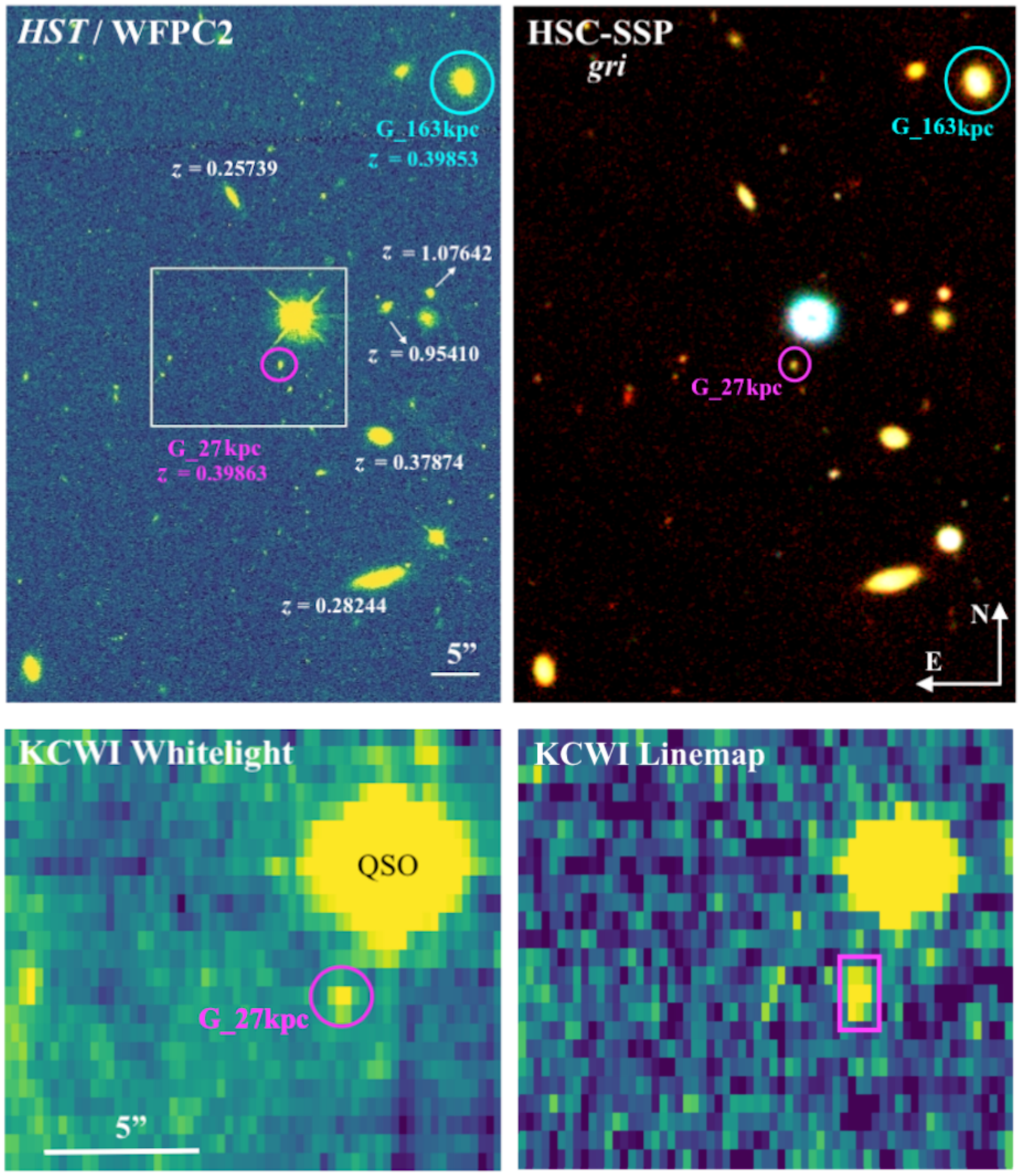}
            \caption{An overview of the Q0122$-$003 field. (top left) {\it HST}/WFPC2 image in the {\it F702W} filter. The galaxies associated with the $z_{\rm abs}=0.3985$ absorber studied here are labeled with cyan (G\_163kpc) and pink (G\_27kpc) circles, whereas other galaxies in the field with spectroscopic redshifts are tagged with their $z_{\rm gal}$ values. The low mass galaxy is $5\farcs3$ southeast of the quasar (QSO), whereas the massive galaxy previously found by \citetalias{muzahid15} is $30\farcs3$ northwest of the quasar sightline. The KCWI field of view is plotted as the white rectangle. (top right) Subaru/HSC image combined with $g,r,i$ bands centred at the quasar. (bottom left) KCWI whitelight image where the spectral direction is summed over a wavelength range of $3500-5500$~{\AA} for each spaxel. (bottom right) KCWI linemap image summed up over $5205-5222$~{\AA} ($v\pm500$~{\kms} around $z_{\rm abs}$). The emission features in this panel include only the quasar and G\_27kpc. Our analysis shows that there are no emission-line galaxies emission at this redshift on top of the quasar. The spaxels used to extract the spectrum of G\_27kpc are boxed with a pink rectangle.} 
    \label{HSC_image}
\end{figure*} 

\subsection{Quasar field imaging, morphology and photometry} 
\label{qso imaging and morphology}

The Q0122$-$003 field was imaged using {\it HST}/WFPC2 and the {\it F702W} filter for a 2100 second exposure (program ID:6619). The WFPC2 Associations Science Products Pipeline (WASPP) was used to reduce and calibrate the image. The magnitude limit of the image is $m_{F702W}\sim26$, which translates to a luminosity limit of $L = 0.002L_{\star}$ and $M_{B}=-14.5$ at $z=0.4$. The {\it HST} image of the Q0122$-$003 field is shown in upper left panel of Fig.~\ref{HSC_image}.

The {\it HST} image was used to model the galaxy orientations following the methods described by \cite{Glenn2011augustGIM2D} and \cite{Glenn2015dec}. This technique quantifies the orientation and morphological parameters of galaxies by fitting two-component disk+bulge models implemented in the GIM2D software \citep{Simard2002}. The bulge and the disk components have a Sersic profile (with $0.2<n <4.0$) and an exponential profile, respectively. 

We also utilized imaging from the Hyper Suprime-Cam Subaru Strategic Program \citep[HSC-SSP;][]{Aihara2018,Aihara2019}, which consists of Wide, Deep and UltraDeep layers. The Q0122$-$003 field is located in the Wide layer, which covers 1400 deg$^{2}$ of the sky at a depth of $r\sim26$. These reasonably deep $grizy$ images, with $5\sigma$ depths of $g\sim26.2$, $r\sim26.6$, and $i\sim26.2$, allow us to detect very faint galaxies within the field \citep{Bosch17}. We show a $gri$-colour image in the upper right panel of Fig.~\ref{HSC_image}. 
We selected CModel magnitudes to obtain the best measure of the total object magnitude in each photometric band. This is because CModel magnitudes are computed from an algorithm that fits galaxy flux distributions and light profiles using both a de Vaucouleurs and an exponential component while accounting for the point spread function (PSF). We further corrected these magnitudes for Galactic dust attenuation by applying Galactic extinction corrections using a two-dimensional dust reddening map produced by the Planck mission collaboration \citep{plank11,planckdust14} and assuming a \cite{cardeli89} attenuation law.

\subsection{Keck/KCWI -- Observations and Reduction}

The Q0122$-$003 field was observed for 960 seconds using the Keck Cosmic Web Imager \citep[KCWI;][]{morrissey18_kcwi} on Keck~II. This observation was conducted under the program ID: $2018$B$\_$W232 on 2018 September 9 UT. The medium slicer with a field of view (FOV) of $16\farcs5 \times 20\farcs4$ along with the BL grating with a central wavelength of 4500~{\AA} and $2\times2$ binning was used. The FOV of the medium image slicer results in a spatial resolution of $0\farcs29\times0\farcs69$, which corresponds to $1.5\times3.7$~kpc at $z=0.4$. The spectral resolving power (R) of the BL grating is $\approx1800$ ($\sim0.625$~{\AA}~pixel$^{-1}$) and covers a wavelength range of approximately 3500$-$5500~{\AA}. The footprint of KCWI is specified with a white rectangle on the {\it HST}/WFPC2 image in the top left panel of Fig.~\ref{HSC_image}.

The data were reduced with the publicly available KCWI Data Reduction Pipeline\footnote{\url{https://github.com/Keck-DataReductionPipelines/KcwiDRP}} using standard settings, but skipping the sky subtraction step. To flux-calibrate the data, we used the standard star kopff27 from the KCWI DRP starlist observed earlier in the night. The data are also vacuum and heliocentric velocity corrected. The standard reduction pipeline introduces a wavelength-dependent gradient perpendicular to the slices and does not adequately remove scattered light from the bright quasar. To correct for these effects, we used the non-sky-subtracted, flux-calibrated cube where the quasar has been masked out. The gradient is divided out in each wavelength bin and slice. A median sky is then determined from the corrected cube and subtracted. Further details will be presented in Nielsen et al.~(in preparation), but this method follows a similar approach to that used in CubEx \citep{Cantalupo19}.
The final reduced datacube has a $3\sigma$ line flux limit of 
$7\times10^{-18}$~erg~s$^{-1}$~cm$^{-2}$, which results in a SFR limit of 0.02 M$_{\odot}$~yr$^{-1}$. These data are shown in the bottom left panel of Fig.~\ref{HSC_image}, where the wavelength dimension has been collapsed between 3500~{\AA} and 5500~{\AA} to produce a whitelight image. 


\subsection{Quasar Spectra -- UV and Optical}

The ultraviolet spectra of background quasar Q0122$-$003 were obtained with our "Multiphase Galaxy Halos" Survey using {\it HST}/COS in Cycle-21 \citep[PID: GO-13398;][]{Glenn2015dec, Glenn2019, muzahid15, muzahid2016, Nikki2017, Pointon2017, pointon19, Pointon2020, MasonNg2019}. The far-UV (FUV) grating G160M was utilized and has a moderate resolving power of $R\sim20,000$, giving a full width at half maximum of $\sim18$~{\kms}, and wavelength coverage of $1410-1780$~{\AA}. The data were acquired from the {\it HST} archive, reduced using the STScI CALCOS V2.21 pipeline \citep{Massa_COS}, and flux-calibrated. We increased the signal-to-noise ratio of our spectrum by co-adding multiple integrations with the IDL code {\sc coadd\_x1d}\footnote{\url{http://casa.colorado.edu/~danforth/science/cos/costools.html}} \citep{Danforth10} and binning by three pixels. This yields a ${\rm S/N}\sim 9-17$ per resolution element. The spectrum was continuum-normalized by fitting line-free regions with smooth low-order polynomials and the wavelengths were vacuum and heliocentric velocity corrected.

The optical spectrum of Q0122$-$003 was obtained with VLT/UVES \citep[][]{Dekker2000UVES} in July 2005 as part of program ID: 075.A$-$0841. The spectrum covers the wavelength range $3290-9466$~{\AA} with a resolution of $R\sim 45,000$. The European Southern Observatory (ESO) pipeline \citep{Dekker2000UVES} was used for the data reduction \citep[for details, see][]{Glenn2011augustGIM2D}. The wavelengths were vacuum and heliocentric corrected for direct comparison to the COS data.

\subsection{Absorption System}
\label{abs_system}
The absorption lines originating from a variety of ionic transitions at $z_{\rm abs} = 0.398$ were specifically studied by \citetalias{muzahid15}, who used VPFIT\footnote{\url{http://www.ast.cam.ac.uk/~rfc/vpfit.html}} to model the absorption features. See \citetalias{muzahid15} for fitting details, along with the measured column densities and limits. 
We summarize their absorption profile fitting and analysis below.


This absorption system is detected in both low ({\CII}, {\NII}, {\MgII}, {\MgI}, {\FeII}, {\SiII} and {\SiIII}) and high ({\CIV}, {\NV} and {\OVI}) ions along with {\Lyb} and {\Lya}\footnote{The higher order Lyman series lines are not covered by the existing COS spectrum.}. A subset of these ions is shown in Fig.~\ref{rotation}. The high resolution UVES spectrum covers the {\MgIIdblt} doublet, which was determined to be best-fit using three Voigt profile components. These components are referred to as L$_{1}$, L$_{2}$ and L$_{3}$ with $v\sim-10,+180,+200$~{\kms}, respectively, with the velocity zero-point set to the redshift of G\_163kpc. All other low ions have similar velocity structure and VP component velocity centroids, which are well-fit using the L$_{1}$, L$_{2}$ and L$_{3}$ components. The {\CII}~$\lambda1036$ and {\SiII}~$\lambda1260$ absorption detected in the COS spectrum resemble the kinematic structures resolved in {\MgII}, although L$_{2}$ and L$_{3}$ are unresolved due to the lower resolution of COS. 


The highly ionized phase is best modeled with the {\OVIdblt} doublet and was fitted with five components, which are referred to as H$_{1}$ through H$_{5}$. This intervening {\OVI} absorber exhibits a large kinematic spread of $\Delta v_{\rm 90}=419$~{\kms} and large column density of $\log N({\OVI})=15.16\pm0.04$, making it one of the strongest intervening {\OVI} absorption systems \citepalias[][and references therein]{muzahid15}. {\NV} is also detected in three components, H$_{3}$, H$_{4}$ and H$_{5}$. {\CIV} is only covered in a low resolution {\it HST}/FOS spectrum and is self-blended so little kinematic information is available, although a lower limit on the column density was determined.  

\citetalias{muzahid15} adopted the line-broadening of {\MgII} and {\OVI} to estimate the maximum allowed column densities of {\HI} in each component of low- and high-ionized gas phase, respectively. The resultant fit profile comprised of two ionization phases is shown in Fig.~\ref{rotation}. The authors implemented photo-ionization modelling for the low and high ionized components separately, using CLOUDY \citep{ferland13_cloudy} to compute the physical properties of the absorbing clouds. 
We summarise the \citetalias{muzahid15} results in Table~\ref{absorption param} for both the low and high ionization phases. For the low ionization phase, the authors found that the metallicity is low for L$_{1}$, L$_{2}$ and L$_{3}$, with lower limits on the metallicity ranging from ${\rm [X/H]}\gtrsim -1.3$ to $-2.0$. 
Although photo-ionization models are well fitted to the data, the metallicity of L$_{2}$ and L$_{3}$ are not well constrained due to the lack of higher order Lyman series lines. However, the metallicity of L$_{1}$ cannot be much higher than ${\rm [X/H]}\gtrsim -2.0$, in order to be consistent with the red wing of the {\HI} absorption.

For the high ionization phase, only the H$_{3}$, H$_{4}$, and H$_{5}$ components were modeled since both {\NV} and {\OVI} are detected and a robust {\HI} can be obtained. The authors found a metallicity of ${\rm [X/H]}\gtrsim 0.3$ for these clouds.
It is worth noting that the metallicity of the high ionization absorption components is $\sim1$~dex greater than the metallicity seen in the low ionization gas. Also note that the $\sim1.5$~dex difference in density between the phases further emphasises the distinct origins of the absorbing gas in each phase.
\citetalias{muzahid15} concluded that both phases are best modelled by photoionized gas, and demonstrated that both collisional ionization equilibrium and non-equilibrium models do not reproduce the observations. 


\section{Results}
\label{result}
In this section, we present the properties of a new galaxy (G\_27kpc) detected at the same redshift as the previously known galaxy (G\_163kpc) at $z_{\rm abs}=0.3985$ along the Q0122$-$003 sightline. G\_27kpc has a line-of-sight velocity separation of $\Delta v=21.43$~{\kms} from the G\_163kpc galaxy previously studied by \citetalias{muzahid15}. Contrary to their results, which attempted to explain CGM properties around an isolated galaxy, the detection of a second galaxy associated with the absorption system leads us to explore the complex nature of the CGM around a pair of galaxies. Table~\ref{galaxies prop} lists the properties of this newly discovered galaxy along with the properties of the previously-studied star-forming galaxy with new measurements of stellar and halo masses using galaxy photometry from present work.     

\subsection{Identification of Dwarf Galaxy G\_27kpc} 

Fig.~\ref{HSC_image} presents the {\it HST}/WFPC2 (top left) and HSC-SSP/Wide $gri$ (top right) images of the Q0122$-$003 field. Galaxies with spectroscopic redshifts regardless of whether they are associated with the absorber studied here are also labeled. The galaxy reported by \citetalias{muzahid15} is labeled as G\_163kpc on the {\it HST} image with a cyan circle, which has been detected at $z_{\rm G\_163kpc}=0.39853\pm0.00003$. This galaxy is $30\farcs39$ away from the quasar line-of-sight with an impact parameter of 163~kpc. The white rectangle on the {\it HST} image marks the KCWI FOV using the medium slicer. In Fig.~\ref{HSC_image} (bottom left), we also show the KCWI whitelight image. The quasar is the brightest object in the field and there are several faint continuum sources in the whitelight image, which also appear in the {\it HST} image.


To identify additional galaxies in this field associated with the $z_{\rm abs}=0.3985$ absorber, we created a line map covering the wavelengths expected for {\OII} emission at this redshift. Fig.~\ref{HSC_image} (bottom right) shows a narrow-band image extracted from the KCWI datacube between $5205\leq\lambda\leq5222$~{\AA}. This wavelength range corresponds to a line-of-sight velocity window of $v\pm500$~{\kms} centred at $z_{\rm abs}$. The quasar is the brightest object in the line map and the only other emission feature is the small galaxy $5\farcs3$ Southeast of the quasar sightline. We have also explored subtracting the quasar continuum within the velocity window of $v\pm500$~{\kms} centred at $z_{\rm abs}$, following methods of \citet{Zabel19_megaflow2}. We do not find any hidden emission from galaxies residing on top of the quasar to within a 3$\sigma$ flux limit of $6.5\times 10^{-18}$~erg~s$^{-1}$~cm$^{-2}$ (SFR$<0.01$~M$_{\odot}$~yr$^{-1}$).  We cannot measure redshifts for any of the other galaxies within the KCWI field due to the short wavelength coverage. Given their red color in the HSC imaging, this is consistent with these other galaxies being at higher redshifts compared to G\_27kpc. Our KCWI data upper wavelength range places these galaxies at $z>0.5$ based on the lack of {\OII} emission. It is less likely that these objects are quiescent galaxies at $z<0.4$ since we do not detect any {\CaII} H \& K absorption. There also remains the possibility of additional galaxies residing outside of the KCWI footprint at a similar redshift of the absorber, which we have not identified. Inspecting the HSC imaging in Fig.~\ref{HSC_image}, we expect that if any additional galaxies were to be identified outside our KCWI field of view, they would also be low mass dwarf galaxies that are further from the quasar sightline than G\_27kpc. This would add to the complexity of this system, however, our main results would remain unchanged.  

\begin{figure}
        \includegraphics[width=\columnwidth]{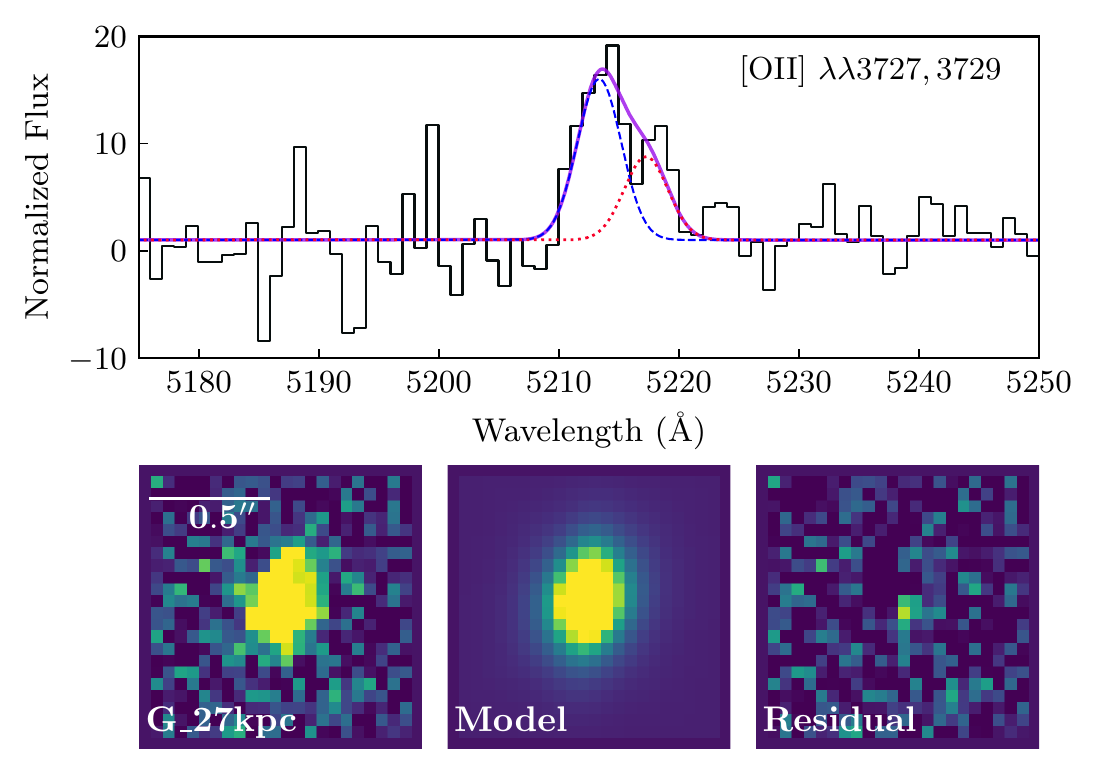}
            \caption{(top) {\OII} doublet emission detected in the KCWI galaxy spectrum with the BL grating summed over the nine spaxels. The black histogram presents the continuum-normalized flux. The resolved {\OII} emission line is modeled with a double Gaussian profile, where blue dashed and red dotted lines represent each {\OII} Gaussian component, whereas pink represents the total fit to the emission line. (bottom left) Zoomed-in {\it HST} image of G\_27kpc. (bottom middle) GIM2D galaxy model. (bottom right) GIM2D residuals. The quasar is located to the upper right of G\_27kpc in these bottom panels, where the quasar sightline probes the galaxy along its projected major axis.}
    \label{galaxyspec}   
\end{figure}

To determine the redshift of the emission-line galaxy, we extracted a summed galaxy spectrum over a total of nine spaxels from the KCWI datacube spatially centred on the emission in the line map (pink rectangle in Fig.~\ref{HSC_image}, bottom right). Fig.~\ref{galaxyspec}(top) presents the galaxy {\OII}~$\lambda\lambda3727,3729$ emission doublet.
We continuum-normalized the spectrum using a quadratic polynomial fit, which excluded the emission line regions. We then fitted a double Gaussian profile to the {\OII} doublet emission by tying both doublet lines to the same redshift and making the assumption that the Gaussian sigma must be at least as large as the spectral resolution ($\sim 71$~km~s$^{-1}$). Fig.~\ref{galaxyspec}(top) shows the total fit from a double Gaussian model in pink, while the blue dashed and red dotted lines correspond to the individual fitted {\OII} doublet emission lines. From our fit we measured the total emission line flux, galaxy redshift and the emission line widths. The spectroscopic redshift of this galaxy is $z_{\rm G\_27kpc}=0.3986\pm0.0001$. This new galaxy is at a similar redshift, within $\Delta v=21.43$~{\kms}, of G\_163kpc. This galaxy, named G\_27kpc, is only $\sim5.3''$ away from the quasar, which corresponds to an impact parameter of 27~kpc. G\_27kpc has apparent magnitudes of $m_{g}=24.19$, $m_{r}=23.26$ and $m_{i}=23.09$ from the HSC-SSP/Wide image. We measure the $B$-band absolute magnitude, $M_{B} = -17.46$, for G\_27kpc by applying a $K$-correction to the $F702W$ apparent magnitude from {\it HST}/WFPC2 observed photometry (see Table~\ref{galaxies prop}) following methods used by \cite{Nikki2013b_magiicat2}. This translates to a $B$-band luminosity of $L_{B}/L^{\ast}_{B} = 0.04$, using $M^{\ast}_{B}$ from \cite{Faber2007}. 


\begin{figure}
        \includegraphics[width=\linewidth]{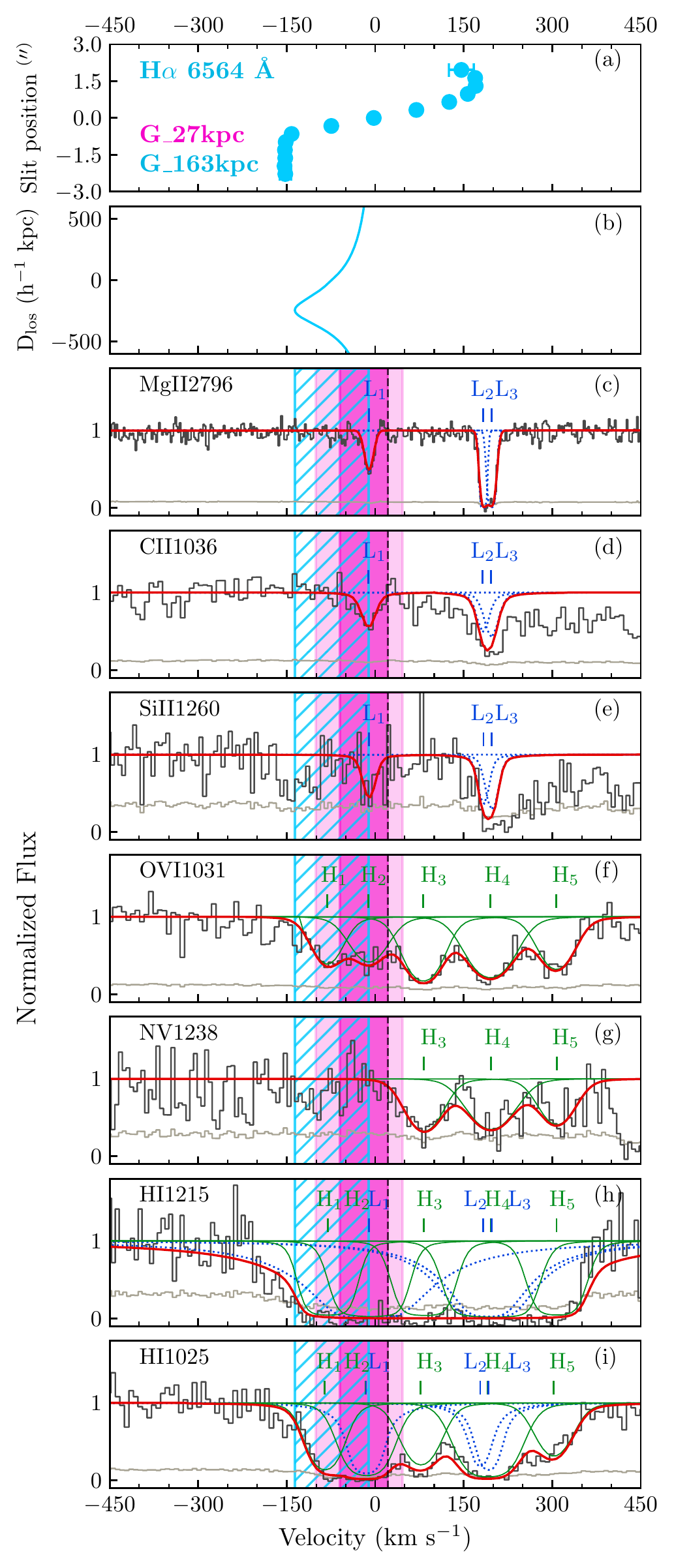}
            \caption{ (a) G\_163kpc rotation curve. (b) Model G\_163kpc disk rotation velocities as a function of distance along the quasar sightline $D_{\rm los}$. (c-i) Absorption profiles and VP fits from \citetalias{muzahid15} for various ions. The G\_163kpc corresponding model velocities are shown as a hatched cyan region in each ion panel. The highlighted dark and light pink regions show the estimated rotation velocity in the direction of the quasar sightline of G\_27kpc.
            The disk model of G\_163kpc is not successful at reproducing the observed absorption velocities, e.g., the blue model and shaded region does not overlap with most of the low and high ionization absorption components. However, our estimated rotation velocity of dwarf galaxy G\_27kpc overlaps with the L$_{1}$ component. 
            The larger velocity range of {\OVI} compared to other ions cannot be explained by the models or envelopes for either galaxy.}
    \label{rotation}
\end{figure} 

\begin{table}
	\centering
		\caption{Absorption line modelled parameters from VPFIT and Photo-ionization Modelling \citepalias{muzahid15}. }
	\label{absorption param}
	\begin{threeparttable}
		\begin{tabular}{ccccc} 
		\hline
		\hline
		VP & $\textgreater$log N$_{\tiny \HI}$\tnote{a} & $\gtrsim$~[X/H]\tnote{a} & log~$U$ & log~$n_{\rm H}$ \\
		Comp & (cm$^{-2}$) & & & (cm$^{-3}$)\\
		\hline
		\hline
		\multicolumn{5}{c}{Low-Ionization Phase}\\
		\hline
		L$_{1}$ & 19.7  &  $-2.00$ & $-3.45$ & $-2.40$ \\
		L$_{2}$ & 19.8  &  $-1.30$ & $-3.25$ & $-2.60$ \\
		L$_{3}$ & 19.7  &  $-1.40$ & $-3.35$ & $-2.50$ \\
		\hline
	    \multicolumn{5}{c}{Mid- to High-Ionization Phases}\\
		\hline
		H$_{3}$ &18.1 & 0.6 & [$-1.70$,~$-1.60$] & [$-4.25$,~$-4.15$]\\
		H$_{4}$ &18.6 & 0.3 & [$-2.00$,~$-1.80$] & [$-4.05$,~$-3.85$]\\
		H$_{5}$ &17.8 & 0.7 & [$-1.60$,~$-1.50$] & [$-4.35$,~$-4.25$]\\
		\hline
	\end{tabular}
	
	\begin{tablenotes}
            \item[a] Lower limits on neutral hydrogen column density and metallicity of each gas phases.
            \item[b] The range of log~$U$ and log~$n_{\rm H}$ in which the photo-ionization model is successful to reproduce the observed column density of {\OVI} and {\NV}.
        \end{tablenotes}
	\end{threeparttable}
\end{table}

\begin{table}

\centering
\setlength{\arrayrulewidth}{0.05mm}
\renewcommand{\arraystretch}{1.5}
    
	\caption{ The measured properties of G\_27kpc and G\_163kpc.}
	\begin{threeparttable}
	
	\begin{tabular}{lccc} 
		\hline
		\hline
		Galaxy properties & G\_27kpc & G\_163kpc  \\
		\hline
		\hline
		RA (J2000) & 01:25:28.96 & 01:25:27.67 \\
		DEC (J2000) & $-$00:06:00.84 & $-$00:05:31.39 \\
		$z_{\rm gal}$ & $0.3986\pm0.0001$ & $0.39853\pm0.00003$\tnote{a} \\
		$\theta~['']$ & 5.36 & 30.39\tnote{a} \\
		$D$~[kpc] & 27.01$\pm$0.02 & 163$\pm$0.1\tnote{a} \\
		$m_g$ & 24.19$\pm$0.04 & 20.864$\pm$0.002 \\
		$m_r$ & 23.36$\pm$0.03 &  19.782$\pm$0.001 \\
		$m_i$ & 23.09$\pm$0.05 & 19.314$\pm$0.002 \\
		$m_z$ & 22.62$\pm$0.07  & 18.954$\pm$0.003 \\
		$m_y$ & 22.58$\pm$0.06 & 18.761$\pm$0.002 \\
		$m_{F702W}$ & 23.52$\pm$0.07 & 19.4\tnote{a} \\ 
		$L_B/L_B^{\ast}$ & 0.04 & 1.4\tnote{b} \\
		$\log(M_{\star}/{\rm M}_\odot)$ & $8.66 \substack{+0.80 \\ -0.10}$ &  $10.50 \substack{+0.57 \\ -0.03}$ \\ 
		$\log(M_{\rm h}/{\rm M}_\odot)$ & $10.34 \substack{+0.79 \\ -0.01}$ & $12.20 \substack{+0.52 \\ -0.08}$ \\
		 $R_{\rm vir}$~[kpc] & $101\substack{+57 \\ -15} $ & $278\substack{+38 \\ -32}~\tnote{a} $\\
		 $D/R_{\rm vir}$ & $0.26$ & $0.59$\\
		 SFR~[M$_{\odot}$~yr$^{-1}$] & $0.08\pm 0.03$ & 6.9\tnote{a}\\
		 $\log({\rm SSFR})$~[yr$^{-1}$] & $-$9.8 & $-$9.7\\
		 $\Sigma_{\rm SFR}$~[M$_{\odot}$~kpc$^{-2}$~yr$^{-1}]$ &0.006  & 0.4\tnote{a} \\
		 $i$~[degrees] & $41.0\substack{+7.3 \\ -8.2}$ & $63.2\substack{+1.7 \\ -2.6}~\tnote{a}$ \\
		 $\Phi$~[degrees] & $1.6 \substack{+34 \\ -1.6}$ & $73.4\substack{+4.7 \\ -4.6}~\tnote{a}$ \\
		 \hline
		 
	\end{tabular}
	
	   \begin{tablenotes}
            \item[a] The measurements are quoted from \citetalias{muzahid15}.
            \item[b] The $B-$band luminosity of G\_163kpc was recalculated and resulted in a larger luminosity than that reported by \citetalias{muzahid15}.
        \end{tablenotes}
	\end{threeparttable}
	\label{galaxies prop}
\end{table}

\subsection{Properties of G\_27kpc and G\_163kpc} 

In the bottom panels of Fig.~\ref{galaxyspec}, we present a zoomed image of G\_27kpc from {\it HST}, the GIM2D galaxy model and the residual to show the quality of the fit. This moderately inclined galaxy is well-resolved, yet the morphology results in a not well-constrained azimuthal angle. The modelled orientation of G\_27kpc yields an inclination angle of $i = 41\substack{+7 \\ -8}$~degrees and an azimuthal angle of $\Phi = 1.6 \substack{+34 \\ -1.6}$~degrees. For G\_163kpc the inclination ($i = 63\substack{+2 \\ -3}$~degrees) and azimuthal angles ($\Phi = 73 \substack{+5 \\ -5}$~degrees) are from \citetalias{muzahid15}. Note that $\Phi = 90$~degrees points the quasar sightline out along the projected minor axis and $\Phi = 0$~degrees points the sightline out along the projected major axis of galaxies. G\_27kpc has its major axis roughly aligned with the quasar sightline, while G\_163kpc has its minor axis aligned with the quasar sightline. The morphological properties for each galaxy are listed in Table~\ref{galaxies prop}.  Therefore, given the different geometries between the two galaxies and the quasar sightline, it is possible that the sightline is probing different mechanisms consistent with one or both galaxies' CGM.


To derive the stellar masses of G\_27kpc and G\_163kpc, we fitted the HSC-SSP five-band CModel photometry using FAST++ \footnote{\url{https://github.com/cschreib/fastpp}}, which is a modification of the spectral energy distribution (SED) fitting code FAST \citep{kriek09}. The code compiles stellar population synthesis (SPS) templates over grids of stellar population parameters and fits SPS models to galaxy flux values and/or one-dimensional spectra to estimate galaxy stellar population properties (stellar mass, age, dust, metallicity, redshift and star formation time scale) by minimizing the $\chi^{2}$. For G\_27kpc we use CModel fluxes that have been corrected for Galactic extinction and the spectrum extracted from the KCWI datacube (plotted in Fig.~\ref{galaxyspec}). For G\_163kpc we use only the CModel fluxes. We used a grid of SPS models from \cite{bc03} and adopted a \cite{chabrier03} initial mass function (IMF) with exponentially declining star formation histories with characteristic timescale, $\tau$, that varies between $10^{7}$~yr and $10^{10}$~yr. Following the \cite{calzetti03} dust extinction law assuming a uniform screen of dust attenuation for the entire galaxy, we let $A_{V}$ vary between $0-4$~mag. We allow the age of the stellar population to vary from $10^{8}$~yr to $10^{10}$~yr. The metallicity is left as a free parameter and varies between $Z = 0.004$~(subsolar), $Z=0.02$~(solar) and $Z = 0.05$~(supersolar). We also fixed the model redshift to the spectroscopic redshift of the galaxies.

The stellar mass estimated by FAST++, with 1$\sigma$ uncertainties, for G\_27kpc is $\log(M_{\star}/{\rm M}_{\odot})={8.7}\substack{+0.8 \\ -0.1}$ and for G\_163kpc is $\log(M_{\star}/{\rm M}_{\odot})=10.50\substack{+0.57 \\ -0.03}$. In order to derive the halo mass, $M_{\rm h}$, we converted the FAST stellar masses to halo mass using the relation reported by \cite{Moster2010}, which accounts for the redshift evolution in the correlation of the measured stellar mass to the virial mass (dark + baryonic matter) of the galaxy. The computed halo masses for G\_27kpc and G\_163kpc are $\log (M_{\rm h} / {\rm M}_{\odot})=10.34\substack{+0.79 \\ -0.01}$, and $\log (M_{\rm h} / {\rm M}_{\odot})=12.20\substack{+0.52 \\ -0.08}$, respectively. 
Note that there is a two magnitude difference between the stellar masses of G\_27kpc and G\_163kpc.
The stellar mass of G\_27kpc classifies it as a dwarf galaxy \citep[e.g.,][]{cos-dwarf, calabro17_dwarf}.  The virial radius of G\_27kpc is $R_{\rm vir} = 101\substack{+57 \\ -15}$~kpc, and G\_163kpc is $R_{\rm vir} = 278\substack{+38 \\ -32}$~kpc, both calculated using the formalism of \cite{Bryan1998}. These values result in virial radius normalized impact parameters of $D/R_{\rm vir} = 0.26$ and $D/R_{\rm vir} = 0.59$ for G\_27kpc and G\_163kpc, respectively.

The SED modelling predicts a metallicity of $Z=0.004\substack{+0.034 \\ -0.004}$~Z$_\odot$ for G\_27kpc and a solar metallicity of $Z~=~0.02\substack{+0.02 \\ -0.01}$~Z$_\odot$ for G\_163kpc, which is consistent with the metallicity of G\_163kpc calculated from emission lines \citepalias{muzahid15}. The mass--metallicity relation derived by \cite{calabro17_dwarf} predicts a sub-solar to solar metallicity for dwarf galaxies with the same stellar mass of G\_27kpc, which is consistent with our modelled metallicity.

We used the {\OII} emission doublet to compute the SFR of G\_27kpc using the relation from \cite{kewley04}. The {\OII} emission-line luminosity, which is not corrected for the interstellar dust reddening of this galaxy, is $\log(L_{{\OII}})=40.28$~erg~s$^{-1}$. 
Since the Chabrier initial mass function \citep{chabrier03} is adopted for fitting the SED of the galaxies, we also changed the amplitude of the \citeauthor{kewley04} relation from a Salpeter IMF \citep{salpeter1995} to a Chabrier IMF \citep{chabrier03} to result in the following equation:
\begin{equation}
\begin{aligned}
\operatorname{SFR}\left(\left[\mathrm{O}{\mathrm{Il}}\right]\right)\left(\mathrm{M}_{\odot}~\mathrm{yr}^{-1}\right)= 4.0 
& \times 10^{-42} L\left(\left[\mathrm{O}{\mathrm{II}}\right]\right)\left(\mathrm{erg}~\mathrm{s}^{-1}\right).
\end{aligned}
\label{sfr}
\end{equation}
We estimated the star formation rate (SFR) of G\_27kpc to be ${\rm SFR}=0.08\pm0.03$~M$_{\odot}$~yr$^{-1}$. G\_163kpc has an {\Ha}-derived ${\rm SFR}=6.9$~M$_{\odot}$~yr$^{-1}$ \citepalias{muzahid15}. The specific star-formation rates (SSFR) for G\_27kpc and G\_163kpc are $\log({\rm SSFR})=-9.8$~yr$^{-1}$ and $\log({\rm SSFR})=-9.7$~yr$^{-1}$, respectively. While the SSFR of G\_163kpc places it on the star-formation main sequence, this is not the case for G\_27kpc. G\_27kpc is roughly one magnitude below the star-formation main sequence when compared to galaxies of similar mass and redshift \citep{calabro17_dwarf}. 

We derive the star formation rate per unit area for G\_27kpc using the half light radius of $R_{\rm h}=1.4$~kpc obtained from the GIM2D modeling discussed in Section~\ref{qso imaging and morphology} (also see the bottom panels of Figure~\ref{galaxyspec}). From the measured SFR surface density, $\Sigma_{\rm SFR}=0.006$~M$_{\odot}$~kpc$^{-2}$~yr$^{-1}$, we do not expect a significant amount of ionized outflowing gas from this dwarf galaxy
\citep[outflows are typically found for $\Sigma_{\rm SFR}>0.1$~M$_{\odot}$~kpc$^{-2}$~yr$^{-1}$, e.g.,][]{Heckman11,sharma17}. On the other hand, G\_163kpc has a star formation rate per unit area of $\Sigma_{\rm SFR}=0.4$~M$_{\odot}$~kpc$^{-2}$~yr$^{-1}$, which is a factor of four above the threshold expected for star-formation driven outflows. 


\subsection{Galaxy--CGM Kinematics} 

We next explore the kinematic relationship between the absorption and galaxies G\_27kpc and G\_163kpc.
In Fig.~\ref{rotation}(a), we show the rotation curve of G\_163kpc obtained from the {\Ha} emission line using the methods of \cite{Glenn2010}. G\_163kpc's rotation curve flattens to a maximum rotation velocity of $\simeq180$~{\kms}. We used a co-rotating thick-disk model from \citet{Steidel2002} to see if any of the absorption could be explained by co-rotation/accretion. This model is dependent on galaxy orientation with respect to the quasar (i.e., $i$ and $\Phi$), the impact parameter, and the rotation velocity, which then predicts the expected rotation velocities through the halo. As shown in Fig.~\ref{rotation}(b), the model predicts a range of plausible gas velocities of $-150\leq V\leq -25$~{\kms} along the line-of-sight ($D_{\rm los}$) for the observed properties of G\_163kpc. In Fig.~\ref{rotation}(c), the shaded cyan region shows the kinematic model velocity range, which does not overlap fully with any of the {\MgII} components, which have velocity centroids at $-$10, 183 and 197~{\kms} and this applies to the low ionization gas kinematics traced by {\CII} (Fig.~\ref{rotation}(d)) and {\SiII} (Fig.~\ref{rotation}(e)). This implies that most of the low ionization gas phase is inconsistent with disk rotation from G\_163kpc. From Fig.~\ref{rotation}(f), the model only overlaps with H$_{1}$ and partially the H$_{2}$ components of {\OVI} that are extended over the velocities of $\sim-90$ to $-20$~{\kms}. The model does not overlap with components H$_3$ through H$_5$, which are the only components in which {\NV} (Fig.~\ref{rotation}(g)) is detected.

The spatial resolution of the KCWI data does not allow us to directly measure a fully resolved rotation curve for G\_27kpc. However, we are able to constrain the direction of rotation using the moderately spatially-resolved galaxy spectrum. We compute the direction of rotation by comparing the emission-line centroid from the spectrum of the middle three spaxels to the  emission-line centroid from the spectrum of the bottom three spaxels (the top three spaxels do not have enough signal) highlighted by the pink box in Fig.~\ref{HSC_image}. 
We compute a velocity difference of $55\pm45$~{\kms}, and although the error is large, the sign of the value suggests that the major axis pointing away from the quasar is redshifted while the major axis pointed towards the quasar is blueshifted.
Given that G\_27kpc is not resolved enough to compute a full rotation curve, we examine the velocity dispersion of its ISM. The measured {\OII} velocity dispersion is $\sigma_{\OII}=69\pm28$~{\kms}, which was obtained by subtracting the instrumental dispersion $\sigma_{\rm instrument}=70.73\pm0.008$~{\kms} from the observed dispersion of the {\OII} modelled emission profile. This translates to an emission-line FWHM of $\sim 162$~{\kms} for G\_27kpc. If we assume that the bulk of the ISM kinematics measured here is from galaxy rotation and not from random motion, then we can compare our FWHM with the Tully--Fisher relationship. \cite{conselice2005} studied a sample of 101 disk-galaxies at $0.2<z<1.2$ to determine the luminosity and stellar mass Tully--Fisher relation and the authors showed a correlation between $M_{\star}$ and the maximum rotational velocity of galaxies. The approximate rotation velocity of G\_27kpc would then be equal to ${\rm FWHM}/2\sim81$~{\kms}, which is in agreement with galaxies having the same stellar mass for $z<0.7$ from \cite{conselice2005}.


The shaded regions of Fig.~\ref{rotation}(c-i) show the estimated rotation velocity in the direction of the quasar sightline of G\_27kpc (dark pink) and the quadratic sum of the redshift error and the rotation velocity error (light pink). As previously stated, we assume that the kinematics of the {\OII} emission lines are dominated by rotation. We find that L$_{1}$ is consistent with the kinematics of G\_27kpc and inconsistent with the direction of L$_{2}$ and L$_{3}$. G\_27kpc and G\_163kpc have velocities consistent with H$_{1}$ and H$_{2}$ and inconsistent with bulk of the high ionization phase. It is clear though that both galaxies have kinematics that are consistent with some of the absorption, yet neither can explain all of the absorption with rotation alone when including the kinematics of both galaxies.

\vspace{0.2in}

In summary, while both galaxies are at nearly the same redshift and could host the detected multi-phase CGM absorption, they are very different objects. While the G\_163kpc galaxy reported by \citetalias{muzahid15} is a massive, star-forming, and likely outflowing galaxy that is at a distance of $0.59R_{\rm vir}$ from the quasar sightline, G\_27kpc is a dwarf with low SFR and SSFR and resides at a distance of $0.26R_{\rm vir}$ from the quasar sightline. In the next section, we discuss the possible origins of the CGM.

\section{Discussion}
\label{sec:discussion}
We have discovered a new dwarf galaxy, G\_27kpc, at a similar redshift as a previously-known galaxy, G\_163kpc, that was thought to be the only host of multi-phase absorption detected in the spectrum of a background quasar. \citetalias{muzahid15} previously concluded that the absorption arose from outflows (high ionization phase) and recycled gas (low ionization phase) from G\_163kpc. Given the discovery of our new galaxy, we discuss how the previous interpretation may change provided this new information.

\subsection{Low-Ionization Phase Gas}

The quasar sightline intersects a $W_r({\MgII})=0.40 \pm 0.01$~{\AA} absorption system, which is 27~kpc ($0.26R_{\rm vir}$) away from a dwarf and 163~kpc ($0.59R_{\rm vir}$) away from a massive star-forming galaxy. \cite{Nikki2013b_magiicat2} found a significant ($\sim 8\sigma$) anti-correlation between the rest-frame {\MgII} equivalent width and the impact parameter of 182 absorber--galaxy pairs. Using their sample, we find that 87 percent of all absorbers (83/95) with an equivalent width of $W_r({\MgII})\geq0.4$~{\AA} reside within $D/R_{\rm vir}\leq 0.3$, while 98 percent of these absorbers (93/95) reside within $D/R_{\rm vir}\leq 0.6$. Based on statistics of a large sample, there is only a 2 percent chance that the {\MgII} absorption is associated with G\_163kpc. It is much more probable that the low ionization phase traced by {\MgII} is associated with the dwarf galaxy G\_27kpc. 


 We now focus our discussion on the likely origins of the L$_{1}$ cloud. L$_{1}$ has the lowest metallicity limit of the low ionization components with ${\rm [X/H]}\gtrsim-2.0$, which is likely not much higher than the number quoted from \citetalias{muzahid15}. The metallicity of L$_{1}$ is comparable to metal-poor IGM gas \citep[e.g.,][]{Carswell2002,Shull2014}. It is also within the low metallicity peak (${\rm [X/H]}\sim -1.8$) of the bimodal metallicity distribution of \citet{Lehner2013} and \citet{Wotta16}, which is assumed to originate from gas accretion \citep[][]{Hafen19}. 
 
 In Figure~\ref{rotation}, we have shown that L$_{1}$ does not align with the rotation of G\_163kpc, where the quasar sightline is located along its projected minor axis. On the other hand Figure~\ref{rotation} shows that L$_{1}$ is consistent with the rotation direction and velocity dispersion of G\_27kpc, where the quasar sightline is located along the projected major axis of G\_27kpc ($ \Phi = 1.6\substack{+34 \\ -1.6}$~degrees; see Fig.~\ref{galaxyspec}). We further determined that L$_{1}$ is consistent with the rotation direction and velocity dispersion of G\_27kpc. This is consistent with studies that have shown {\MgII} gas along the projected galaxy major axis is commonly found to co-rotate with galaxies and is consistent with gas accretion \citep[see review by][]{Glenn2017}. Therefore, we infer that L$_1$ could be gas that is accreting onto G\_27kpc and not associated with G\_163kpc.
 
 We now focus our discussion on the likely origins of the L2 and L3 clouds. As shown in Figure~\ref{rotation}, these two higher velocity components of the low ionization phase cannot be explained by the rotational velocities of G\_27kpc and G\_163kpc.
 Furthermore, the metallicities of L$_2$ and L$_3$ are less well constrained and have limits of [X/H]~$\gtrsim-1.4$ \citepalias[for details, see][]{muzahid15}. However, \citet{Hafen19} showed that satellite winds from dwarf galaxies have a range in metallicity of ${\rm [X/H]}=-2$ to $-0.5$, which is likely consistent with the metallicity of L$_2$ and L$_3$. We note that G\_27kpc currently has a low star-formation rate (${\rm SFR} = 0.08\pm 0.03$~M$_{\odot}$~yr$^{-1}$) and star-formation rate surface density ($\Sigma_{\rm SFR} = 0.006$~M$_{\odot}$~yr$^{-1}$~kpc$^{-2}$). While G\_27kpc is not expected to have strong outflows at the present time, these outflows could have occurred in the distant past given the recycling time for outflows is on the order of a billion years \citep{Oppenhimer08}. \cite{Firesim2017} showed that intergalactic/wind transfers from satellites to the central galaxy is a dominant CGM process at lower redshifts. Although the gas is redshifted relative to both galaxies, \cite{Firesim2017} shows that the gas trajectories can be quite varied due to the large range of galaxy--galaxy orientations happening over a long timescale. It is likely that the L$_2$ and L$_3$ components detected at higher velocities may then trace previously outflowing gas from G\_27kpc that is transferring to G\_163kpc and is part of the intragroup medium. 


Both the L$_{2}$ and L$_{3}$ components are offset by an average line-of-sight velocity of $v_{\tiny \rm LOS}\sim 170$~{\kms} from G\_27kpc. The line-of-sight velocity represents a lower limit on the total velocity of the gas. The virial velocity of G\_27kpc is $v_{\rm vir}=79 \substack{+88 \\ -29}$~{\kms} and the escape velocity at $D/R_{\rm vir}=0.3$ is $v_{\rm esc} = 101\substack{+114 \\ -3}$~{\kms}. The absorption velocity offset of L$_{2}$ and L$_{3}$ are larger than the virial velocity and larger than the escape velocity (although they are lower than the upper bound on $v_{\rm esc}$ due to a larger uncertainty on the mass).  It remains plausible then that the gas in these two components could be escaping the G\_27kpc halo via ancient outflows. This is consistent with the findings of \citet{schroeter_19_megaflow3}, who have shown that outflow velocities only exceed the escape velocity in galaxies with $\log(M_{\star}/{\rm M}_{\odot}) < 9.6$, and consistent with the idea that this gas is transferring from G\_27kpc to G\_163kpc.


Overall, we infer that part of the cool CGM (L$_{1}$) is likely accreting onto G\_27kpc given the consistency between the galaxy and L$_{1}$ CGM kinematics, along with the lower metallicity of L$_{1}$. We further infer that the remaining cool CGM (L$_{2}$ and L$_{3}$) is consistent with ancient star-formation-driven outflows originating and escaping from G\_27kpc. These ancient outflows are likely being transferred to the more massive G\_163kpc since they have velocities larger than the escape velocity of G\_27kpc.


  \subsection{Mid- to High-Ionization Phases}
  
  The high ionization phase of the absorption system is quite different from the low ionization phase. The metallicity derived for three of the well-constrained components, H$_{3}$, H$_{4}$, and H$_{5}$, is super-solar at ${\rm [X/H]}\geq0.3$.  Here we discuss the possible origins of this gas phase.
  
 \cite{chen2001_report} reported strong {\CIV} $\lambda\lambda1548,1550$ observed by {\it HST}/FOS in the Q0122$-$003 field. It has a column density of $\log N({\CIV})\geq14.7$~{\cmsq}, with an unresolved velocity structure. \citetalias{muzahid15} noted that their models produced a significant amount of {\CIV} only in the high-ionization phase for this absorber while no {\CIV} could exist in the low ionization phase. More generally, \citet{Burchett2016} found that {\CIV} is preferentially associated with $M_{\star}> 10^{9.5}$~M$_{\odot}$ galaxies, while lower-mass galaxies rarely exhibit significant {\CIV} absorption with a covering fraction of only $9\substack{+12 \\ -6}$~percent. Therefore these results would suggest that little-to-no {\CIV} should be associated with low mass galaxies like G\_27kpc ($M_{\star} = 10^{8.7}$~M$_{\odot}$) but having  {\CIV} detected in high mass galaxies like G\_163kpc ($M_{\star} = 10^{10.5}$~M$_{\odot}$) is more common. \cite{Burchett2016} further found that $\sim57$~percent of galaxies residing in low-density environments exhibit the {\CIV} absorption, while none of the galaxies in denser regions ($N_{\rm gal}\geq7$) have detected {\CIV}. This also suggests that our absorption system is associated with a low density pair of galaxies, with one dominating the mass distribution.
 
\begin{figure}
        \includegraphics[width=\linewidth]{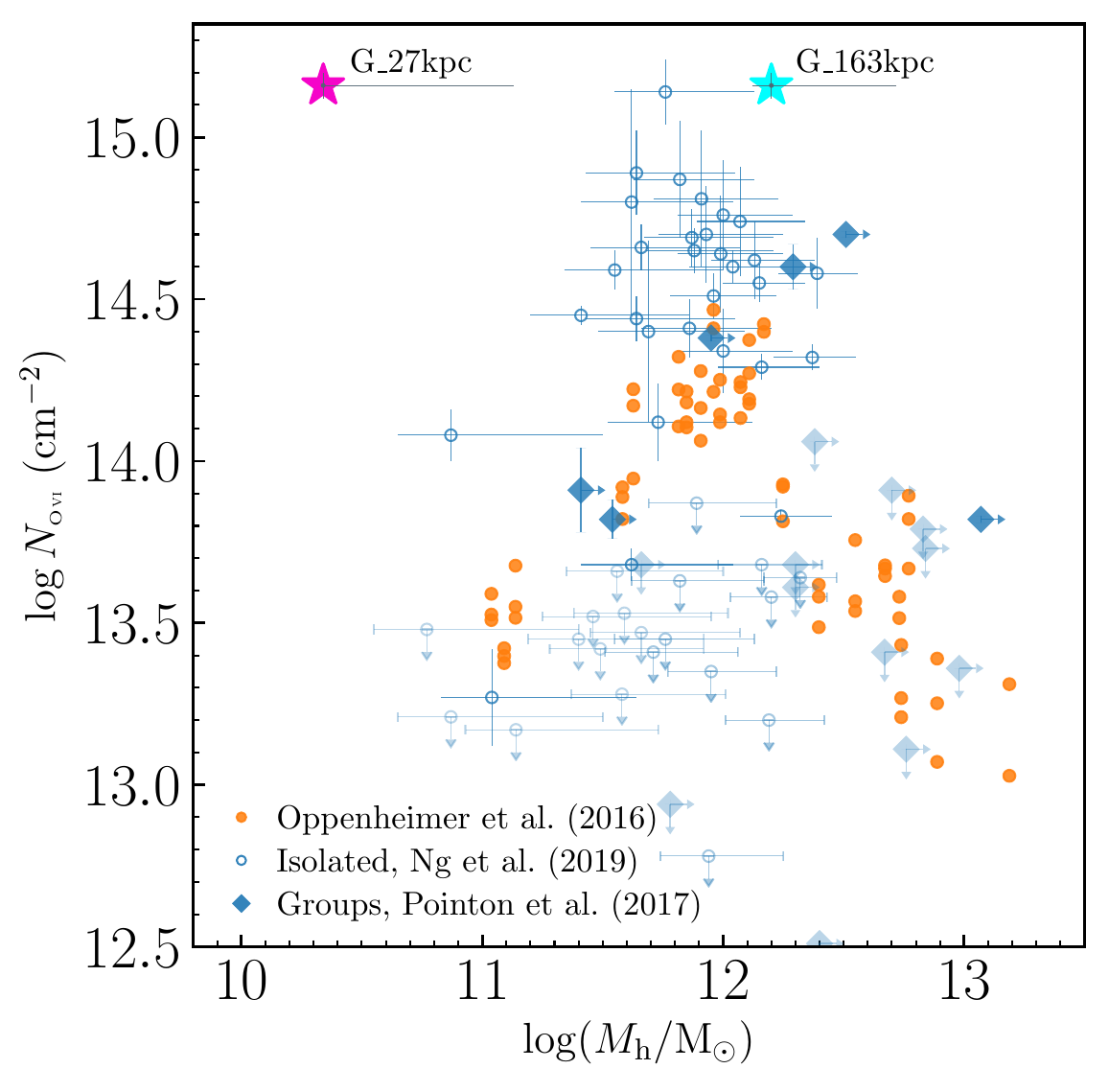}\hfill
            \caption{The distribution of {\OVI} column density as a function of galaxy halo mass studied by
            \citet{oppenheimer16} and \citet{MasonNg2019}. 
            Orange points are the simulated galaxies, larger solid blue diamonds represent the group environments, and the smaller open blue diamonds refer to isolated galaxies. The group galaxies have lower limits on mass. The trend shows that the column density of {\OVI} is greatest for galaxies of $M_{\rm h}\sim 10^{12}$~M$_{\odot}$. The G\_27kpc and G\_163kpc are highlighted with pink and cyan stars, respectively. Even though both galaxies are plotted at the same column density, the massive halo of G\_163kpc resides around the peak of this trend and most probably contributes the {\OVI} components detected in the absorption system. The mass of G\_27kpc is too small to contribute a significant amount of {\OVI} column density.}
    \label{OVI plot}
\end{figure}

The highly ionized absorption observed here has a distinct kinematic structure and a very strong absorption profile spanning a wide range of velocities ($\Delta v_{\tiny \OVI} = 419$~{\kms}, $\Delta v_{\tiny \NV} = 285$~{\kms}), which is in high contrast to the low-ionization phase. \cite{Pointon2017} compared the {\OVI} absorption from group ($N_{\rm gal}=2-8$) and isolated environments. They found that compared to isolated systems, group galaxies are associated with absorption that has smaller average equivalent widths. Using a pixel-velocity two-point correlation function approach, they concluded that the {\OVI} absorption detected in group environments exhibits a significantly narrower velocity spread compared to isolated galaxies. They also argued that the warm/hot CGM does not arise from a superposition of halos within the group but from the intragroup medium. Our two galaxies have similar properties to the groups found in \cite{Pointon2017}, yet the {\OVI} and {\NV} have high column densities with a broad velocity spread. This is in contrast to \citeauthor{Pointon2017}, as we should expect low column densities and velocity spreads. Therefore, the high ionization gas detected here could not be associated to an intragroup environment. Instead, it is highly likely that the high ionization gas is only associated with one of the galaxies.

We compare the estimated ISM metallicities to the absorption line metallcities. G\_163kpc has an ISM with solar metallicity of $12 + \log({\rm O/H}) = 8.68\pm0.02$ (${\rm [O/H]}=-0.01$; \citetalias{muzahid15}), while we expect a lower metallicity for G\_27kpc given its lower stellar mass. The high ionization phase CGM metallicity computed from photo-ionization modelling yields ${\rm [X/H]}\gtrsim0.3$. Only the solar metallicity G\_163kpc would be able to produce such high metallicities in star formation-driven outflows and these metallicities could not arise from a small metal-poor dwarf galaxy. This suggests that the high ionization gas can be completely attributed to G\_163kpc.

The {\OVI} absorption is kinematically spread over $\Delta v_{90}=419$~{\kms}, which is among the four systems with the largest kinematic spreads at $z<1$ \citepalias[][and references therein]{muzahid15}. \cite{MasonNg2019} found that {\OVI} absorption along the minor axis of edge-on galaxies has a large optical depth and the absorption velocity spread spans the systemic velocity, which could be related to bipolar outflows. G\_163kpc has an inclination of $i=63$~degrees, which is moderately inclined, but the large velocity spread found in the high ionization phase is consistent with the expectation of outflows originating from a galaxy that has a large SFR surface density of $\Sigma_{\rm SFR} =0.4$~M$_{\odot}$~yr$^{-1}$~kpc$^{-2}$. It is also clear from Fig.~\ref{rotation} that the majority of the {\OVI} cannot be explained from a co-rotation/accretion scenario for G\_163kpc, given that expected rotation velocities are only consistent with H$_{1}$ and not with the vast majority of the {\OVI} or any of the {\NV}. In addition, accreting gas would not be expected to have such high metallicity. 
This provides further supporting evidence that the high ionization phase arises from G\_163kpc.



Using cosmological simulations, \cite{oppenheimer16} reported that although the total oxygen abundance increases with halo mass, the {\OVI} follows a different distribution (see the orange points in Fig.~\ref{OVI plot}). Galaxies also exhibit a higher fraction of collisionally-ionized {\OVI} relative to photo-ionized {\OVI} with increasing halo mass, although even the most massive galaxies can have a significant amount of photo-ionized {\OVI}. The authors determined that the {\OVI} column density is sensitive to the virial temperature of the halo whereby {\OVI} peaks in $L_{\star}$ galaxies where the virial temperature is ideal for {\OVI} ionization, and becomes lower for lower and higher mass systems where the halo temperatures are less optimal. \cite{MasonNg2019} examined the dependence of 31 {\OVI} absorbers (Fig.~\ref{OVI plot} open and closed blue diamonds, representing isolated galaxies and group environments, respectively) and found that their column density and mass distribution is consistent with the predictions from the simulations as shown in Fig.~\ref{OVI plot}. We show the locations of G\_163kpc and G\_27kpc on  Fig.~\ref{OVI plot}, where we plot both galaxies at the total $\log N_{\tiny \OVI}$ of the absorption system, neglecting that the gas may be physically partitioned between galaxies. Note that while G\_163kpc sits near the peak of the halo mass--$N_{\tiny \OVI}$ relation, G\_27kpc is located at the very low mass end of the plot. It is clear that G\_27kpc is expected to not have any significant {\OVI} absorption within its halo, while G\_163kpc can comfortably accommodate all of the {\OVI} absorption detected here. If the {\OVI} belongs to the group halo encompassing both galaxies, then one would expect less {\OVI} than we currently detect \citep{Pointon2017}. Therefore, this further supports the idea that the high ionization phase arises solely within the halo of G\_163kpc and very little-to-no contribution arises from G\_27kpc.

Overall, we conclude that the high ionization phase arises only within the halo of the massive G\_163kpc galaxy. The large kinematic spread of the gas, the quasar probing the galaxy minor axis, the high CGM and ISM metallicities, and the high $\Sigma_{\rm SFR}$ of this galaxy strongly suggests that the high ionization phase arises from outflows.

\section{Conclusions}
\label{sec:conclusion}
We re-examined the Q0122$-$003 field previously studied by \citetalias{muzahid15}, who found a massive $M_{\star}=10^{10.5}$~M$_{\odot}$ star-forming (${\rm SFR}=6.9$~M$_{\odot}$~yr$^{-1}$) galaxy (G\_163kpc) associated with outflowing gas ($\Sigma_{\rm SFR}=0.4$~M$_{\odot}$~kpc$^{-2}$~yr$^{-1}$) residing at 163~kpc along the galaxy's minor axis. The absorption system is multi-phase and complex, with the low- and high-ionization phases displaying very different kinematics. We have obtained new KCWI data, where we have identified a new dwarf galaxy (G\_27kpc) at the same redshift ($\Delta v=21.43$~{\kms}), which is only 27~kpc away from the quasar sightline. We have found:

\begin{itemize}
    \item The new dwarf galaxy has a mass of $M_{\star}=10^{8.7}$~M$_{\odot}$ with an {\OII} velocity dispersion of $\sim81$~\kms. We assume that the velocity dispersion is rotation-dominated, which then results in a rotation speed consistent with those found for similar dwarf galaxies. 
    
    \item G\_27kpc has a star formation rate of $\rm SFR=0.08$~M$_{\odot}$~yr$^{-1}$ and a specific star formation rate of $\rm SSFR=10^{-9.8}$~yr$^{-1}$. Given these values, it is located slightly below the star-formation main sequence defined for dwarf galaxies. It also has a star-formation rate surface density of $\Sigma_{\rm SFR}=0.006$~M$_{\odot}$~kpc$^{-2}$~yr$^{-1}$, which indicates that it does not currently drive strong outflows. 

     \item G\_27kpc has its major axis pointed towards the quasar sightline ($\Phi= 1.6$~degrees) and it is a moderately inclined galaxy ($i=41.0$~degrees). The quasar sightline is located within $0.26R_{\rm vir}$ of the galaxy. 

\end{itemize}

From our new census of galaxies in this field, we have revised the \citetalias{muzahid15} conclusions regarding the origins of the absorption-line system, which was previously thought to all originate from massive outflows from the distant galaxy G\_163kpc. 
We infer the following about the low ionization phase:

\begin{itemize}
  
    \item The virial radius-normalized impact parameter for G\_163kpc ($D=163$~kpc, $D/R_{\rm vir}=0.59$) is inconsistent with the expectation of the $W_r=$0.4~{\AA} {\MgII} absorption system along the quasar sightline (less than 2 percent). On the other hand, G\_27kpc is at a virial radius-normalized impact parameter ($D=27$~kpc, $D/R_{\rm vir}=0.26$) where {\MgII} is seven times more likely to occur at this distance.
    
    \item Mapping the kinematics of the absorption components shows that L$_{1}$ ($v\sim-10$~{\kms}), which has a metallicity similar to IGM filaments at this redshift (${\rm [X/H]}\gtrsim-2.0$), overlaps with the likely rotation velocity of G\_27kpc ($\pm81$~{\kms}). This suggests that L$_1$ is likely accreting onto G\_27kpc. However, the other two components (L$_{2}$ and L$_{3}$) reside at higher velocities ($\sim+190~{\kms}$), even higher than the escape velocity of G\_27kpc (101~{\kms}), and are not consistent with accreting gas for G\_27kpc. Instead this suggests that the gas traced by these components is escaping the G\_27kpc halo. None of the components are consistent with the rotation direction of G\_163kpc ($-150\leq v\leq-25$~{\kms}).
    
     \item The metallicity limits of L$_{2}$ and L$_{3}$ (${\rm [X/H]}\gtrsim-1.4$) and the results from \citet{Hafen19} likely rules out the possibility of this gas being accretion from IGM. A possibility is that these components are satellite winds due to intergalactic transfers from the dwarf G\_27kpc to the massive galaxy G\_163kpc, consistent with the observed kinematics. 
\end{itemize}

We infer the following about the mid-high ionization phase:

\begin{itemize}
    \item G\_163kpc has a halo mass at the peak of the {\OVI} ionization in galaxies. In contrast, G\_27kpc has a halo mass $100$ times smaller than G\_163kpc and would not be expected to have a significant amount of {\OVI}. Given the fact that the velocity spread of the {\OVI} absorption line ($\Delta v_{90} = 419$~\kms) is one of the largest known at $z<1$, this suggests that G\_163kpc is the likely host of the majority of the {\OVI}.
    
    \item The presence of strong {\NV} and {\CIV} are inconsistent with the small halo mass of G\_27kpc, which lead us to infer that this phase is likely fully associated with G\_163kpc and not related to a group environment. {\CIV} is further inconsistent with G\_27kpc because dwarfs are not expected to host {\CIV} absorption \citep[e.g.,][]{Burchett2016}. 
    
    \item The super-solar metallicity of the high-ionization phase gas (${\rm [X/H]}\gtrsim0.3$) could arise from metal-enriched outflows of G\_163kpc, which has a high star-formation rate surface density, a solar ISM metallicity, and likely has strong outflows.  
    
\end{itemize}

Our work shows that in some galaxy environments, the CGM is complex, especially when the low and high ionization phases differ significantly in their kinematics and metallicities. Our results are consistent with not all gas phases arising in the CGM of a single galaxy, but some phases arise in the CGM of one or within the intragroup CGM with material being transferred between the galaxy pair.  We have determined that the mid- to high-ionization phase is solely associated with the more massive galaxy G\_163kpc. This phase likely arises from star-formation-driven outflows for a galaxy with a halo mass that is at the peak of {\OVI} ionization. On the other-hand, the low-ionization phase is inconsistent with originating from G\_163kpc, and is more consistent with one of the components being accreted onto G\_27kpc, while the other two components are consistent with being ancient outflowing gas originating from G\_27kpc and undergoing intergalactic transfer toward G\_163kpc. 

The additional discovery of this dwarf galaxy has modified the conclusions of the previous work and sheds new light on the origins of the gas. It also emphasises that single integrated metallicities may mask the ability to distinguish certain physical mechanisms within the CGM, and thus it is important to perform component-by-component modelling whenever possible. This work further demonstrates the power of integral field spectrographs and shows that other works may be revised once their fields are spectroscopically complete even within 100~kpc.

\section*{Acknowledgements}
We thank the referee for their helpful comments that improved the manuscript. G.G.K.~and N.M.N.~acknowledge the support of the Australian Research Council through {\it Discovery Project} grant DP170103470. Parts of this research were supported by the Australian Research Council Centre of Excellence for All Sky Astrophysics in 3 Dimensions (ASTRO 3D), through project number CE170100013. C.W.C. and J.C.C. were supported by the National Science Foundation through Collaborative Research grant AST-151786 and by NASA through {\it HST} grant GO-13398 from the Space Telescope Science Institute, which is operated by the Association of Universities for Research in Astronomy, Inc., under NASA contract NAS5-26555. Some of the data presented herein were obtained at the W.~M.~Keck Observatory, which is operated as a scientific partnership among the California Institute of Technology, the University of California and the National Aeronautics and Space Administration. The Observatory was made possible by the generous financial support of the W.~M.~Keck Foundation. KCWI and ESI observations were supported by Swinburne Keck programs 2017B\_W270, 2018A\_W185, 2018B\_W232 (KCWI) and 2014A\_W178E (ESI). The authors wish to recognize and acknowledge the very significant cultural role and reverence that the summit of Maunakea has always had within the indigenous Hawaiian community. We are most fortunate to have the opportunity to conduct observations from this mountain.

\section*{Data Availability}
The data underlying this paper will be shared on reasonable request to the corresponding author.


\bibliographystyle{mnras}
\bibliography{paper} 







\bsp	
\label{lastpage}
\end{document}